\documentclass{jaa}

\usepackage{graphicx}

\begin{document}\sloppy


\title{i-process nucleosynthesis: observational evidences from CEMP stars}


\author{Partha Pratim Goswami\textsuperscript{1},  Aruna Goswami\textsuperscript{1}}
\affilOne{\textsuperscript{1}Indian Institute of Astrophysics, Koramangala, Bangalore 560034, India.\\}


\twocolumn[{

\maketitle

\corres{partha.pg@iiap.res.in; aruna@iiap.res.in}

\msinfo{30 August 2020}{05 October 2020}

\begin{abstract}
The surface chemical compositions of a large fraction of 
Carbon-Enhanced Metal-Poor (CEMP) stars, the so-called CEMP-r/s stars,
 are known to exhibit enhancement 
of both s- and r-process elements. For these stars, the 
heavy element abundances cannot be explained either by s-process or 
r-process nucleosynthesis alone, as the  production sites of s- and r-process 
elements  are very  different, and these two processes produce distinct 
abundance patterns. Thus, the observational evidence of the double enhancement 
seen in  CEMP-r/s stars  remains a puzzle as far as the origin of the elements 
is concerned. In this work, we have critically analysed  the observed 
abundances of  heavy elements in a sample of eight CEMP-r/s stars from 
literature, to trace the origin  of the observed  double enhancement. 
Towards this,   we have conducted a
parametric-model based analysis to delineate the  contributions of 
s- and r-process nucleosynthesis to the observed elemental abundances.
We have  further examined  if the i-process (intermediate process) 
nucleosynthesis, that occurs at a high neutron-density
(n ${\sim} 10^{15} cm^{-3}$) produced during proton-ingestion from a H-rich
envelope to the intershell
region of an AGB star, that is  capable of producing both
r- and s-process elements in a single stellar site,  could explain
the observed abundance patterns of the sample stars. Our analysis
shows that the observed abundance patterns of the selected sample of CEMP-r/s
stars could be fairly well reproduced using  the i-process model yields.

\end{abstract}
\keywords{Stars---Nucleosynthesis---i-process.}

}]


\doinum{12.3456/s78910-011-012-3}
\artcitid{\#\#\#\#}
\volnum{000}
\year{0000}
\pgrange{1--}
\setcounter{page}{1}
\lp{1}

\section{Introduction}
\label{sec:Intro}
Most of the elements heavier than iron are produced by slow (`s') and 
rapid (`r') neutron-capture nucleosynthesis processes. In the slow-neutron 
capture process (s-process) the timescale for neutron-capture is usually 
much longer than that for the $\beta$-decay of unstable nuclei along the 
s-process path, whereas, in the rapid neutron-capture process (r-process) 
the timescale for neutron-capture is much shorter than the $\beta$-decay 
process. While the s-process operates with a neutron-density of  
10$^{6-10}$ cm$^{-3}$ (Busso {\em et al.} 1999) in the inter-pulse phases 
of low and intermediate-mass AGB stars (Gallino {\em et al.} 1998), the 
r-process requires very high temperatures and neutron fluxes 
(n $>$ 10$^{20}$ cm$^{-3}$) and is expected to occur during supernova 
explosions and neutron star mergers (Thielemann {\em et al.} 2011; 
Wehmeyer {\em et al.} 2015). These two processes are characterized by 
distinct elemental abundance patterns. The surface chemical compositions of 
a large fraction of Carbon-Enhanced Metal-Poor (CEMP) stars are known to 
exhibit enhancement of s-process elements (CEMP-s stars), a few exhibit 
enhancement in r-process elements (CEMP-r stars), a sizeable fraction shows 
enhancement of both s- and r-process elements (CEMP-r/s stars) and a few 
stars do not  show enhancement of heavy elements (CEMP-no stars) 
(Beers \& Christlieb 2005; Aoki {\em et al.} 2007). In general, CEMP stars are 
characterized by 
[Fe/H]\footnote[1]{\textbf{Notation:} [A/B] = log(N$_{A}$/N$_{B}$)$_{*}$ $-$ log(N$_{A}$/N$_{B}$)$_{\odot}$, where N$_{A}$ and N$_{B}$ are number densities 
of elements A and B respectively.} 
$<$ $-$1.0 and [C/Fe] $>$ 1.0 and   Barium  and Europium are considered as the 
representative elements of s- and r-process respectively 
(Beers \& Christlieb 2005). Chemical composition studies on (CEMP)-r/s 
stars have revealed that the observed heavy element abundances cannot be 
explained either by s-process or r-process nucleosynthesis alone 
(Aoki {\em et al.} 2015, 2017). In order to explain the abundance 
pattern of the CEMP-r/s stars different formation scenarios have been 
proposed, involving different production sites for the s- and r-process 
elements (Jonsell {\em et al.} 2006; Lugaro {\em et al.} 2009; 
Abate {\em et al.} 2016).  All these scenarios are,  however not free from
certain uncertainties, either in explaining  the observed 
frequency of these stars, or the observed abundance patterns. 
An alternative process called `i-process 
(intermediate process) nucleosynthesis' has  recently been suggested 
as a possible production mechanism for CEMP-r/s stars 
(Dardelet {\em et al.} 2014; 
Hampel {\em et al.} 2016; Hampel {\em et al.} 2019).  
Although it has been known for long  (Cowan \& Rose, 1977) that the
proton-ingestion from the convective envelope to the intershell region 
of AGB stars can produce high neutron flux which can initiate i-process 
nucleosynthesis,  only very recently it has been 
explored  to understand the observed abundance patterns of CEMP-r/s stars 
on the basis of i-process
(Dardelet {\em et al.} 2014; Hampel {\em et al.} 2016; 
Hampel {\em et al.} 2019). 
Different sites have been proposed for the proton-ingestion 
episodes (PIE), such as core-helium flash (Fujimoto {\em et al.} 1990; 
Lugaro {\em et al.} 2009), the most massive AGB stars 
(Jones {\em et al.} 2016), the super AGB stars (Doherty {\em et al.} 2015).
However, the physical conditions under which
proton-ingestion can take place  still remain  a matter of debate. 
Many details regarding the site of the i-process  nucleosynthesis that 
operates  with neutron
densities, n $\sim$ 10$^{15}$ cm$^{-3}$ also remain poorly understood.

In this work, in  order to understand the origin of the abundance 
patterns of heavy elements in CEMP-r/s stars,  we have chosen a sample 
of eight stars reported to be CEMP-r/s stars by various authors 
(Goswami {\em et al.} 2006; 
Goswami \& Aoki 2010; Allen {\em et al.} 2012; Hansen {\em et al.} 2019;
Purandardas {\em et al.} 2019) based on their estimates of abundances 
and abundance ratios  of 
heavy elements for these stars.  We have  performed a parametric-model 
based study to delineate the contributions of s- and r-process to the 
observed heavy element abundances. With reference to the sample stars, 
we have discussed and examined  different formation scenarios  to 
understand the processes responsible for the double enhancement seen 
in these stars. We find,  that using  i-process model yields
we  could reproduce the observed abundance patterns of heavy elements
 of the sample of CEMP-r/s stars under this study.

In Section~\ref{sec:previous_study}, we have discussed in brief, 
the sample of CEMP-r/s stars taken from the literature for this study. 
Section~\ref{sec:parametric_model} discusses the procedure and results 
of the parametric-model based study. In Section~\ref{sec:formation_scenarios}, 
different formation scenarios have been discussed in the context of the 
abundance pattern observed  in the sample stars. 
 Section~\ref{sec:i_process} 
discusses  the i-process models and results of the comparison of the model
predictions with the observed heavy-element abundances of our sample stars. 
We have drawn the conclusions in Section~\ref{sec:conclusion}

\section{Sample of CEMP-r/s stars: CD$-$28~1082, CS~29503$-$010, CS~29528$-$028, HD~209621, HE~0002$-$1037, HE~0059$-$6540, HE~0151$-$6007, HE~1305$+$0007}
\label{sec:previous_study}
\paragraph*{}
The objects are selected following CEMP stars criteria (i.e., [Fe/H] $<$ $-1$,
and [C/Fe] $>$ 0.7). In our sample, the CEMP-r/s stars have metallicity
in the range $-$1.70 $<$ [Fe/H] $<$ $-$2.70. In 
Purandardas {\em et al.} (2019) we have  derived the atmospheric parameters 
and, for the first time, estimated the elemental abundances of CD$-$28~1082.
The object was  found  to be a CEMP-r/s star with [Ba/Fe] = 2.09, 
[Eu/Fe] = 2.07 and [Ba/Eu] = 0.02 with  a $^{12}$C/$^{13}$C ratio ${\sim}$ 16.
 The objects HE~0002$-$1037, HE~0059$-$6540 and HE~0151$-$6007 have been 
reported to be CEMP-r/s by Hansen {\em et al.} (2019). These three objects 
exhibit abundances of Ba and Eu in the ranges 1.7 $<$ [Ba/Fe] $<$ 2.3 
and 1.5 $<$ [Eu/Fe] $<$ 2.3 with [Ba/Eu] $<$ 0.5 in each case. 
Hansen {\em et al.} (2019) estimated the $^{12}$C/$^{13}$C ratio to be 24 
and 1 for the objects HE~0002$-$1037 and HE~0059$-$6540 respectively. The 
kinematic analysis shows that all these four sample stars belong to the 
inner halo population in the Galaxy. Allen {\em et al.} (2012) 
 classified the objects CS~29503$-$010 and CS~29528$-$028 
to be CEMP-r/s stars  with  [Ba/Fe] = 1.81 
\& 2.49 and [Eu/Fe] = 1.69 \& 2.16 respectively on the basis of their 
analysis. Elemental abundances of two CEMP-r/s stars HE~1305$+$0007 and 
HD~209621 have been taken from Goswami {\em et al.} (2006) and 
Goswami \& Aoki (2010) respectively. Tsuji {\em et al.} (1991) and 
Goswami {\em et al.} (2006) estimated $^{12}$C/$^{13}$C ratio to be
 $\sim$ 10 for both the  objects HD~209621 and HE~1305$+$0007.  
 
The atmospheric parameters of the sample stars are presented in 
Table~\ref{tab:atm_para}. The abundance ratios of neutron-capture elements
and carbon with respect to Fe  are presented in Table~\ref{tab:abundances}.

{\footnotesize
\begin{table}
\caption{\bf{Atmospheric parameters of the sample stars. }}
\centering
\label{tab:atm_para} 
\begin{tabular}{lccccccc}
\hline
Star name       		& T$_{eff}$  & log g   &$\zeta$        &     [Fe/H]      \\
                		&    (K)     &         & (km s$^{-1}$) &                 \\
                		&            &         &               &                 \\
\hline
CD$-$28~1082{$^{\textit{1}}$}   &  5200      &  1.90   & 1.42          &$-$2.45          \\
CS~29503$-$010{$^{\textit{2}}$} &  6050      &  3.66   & 1.60          &$-$1.70          \\
CS~29528$-$028{$^{\textit{2}}$} &  7100      &  4.27   & 1.20          &$-$2.15          \\
HD~209621{$^{\textit{3}}$}      &  4500      &  2.00   & 2.00          &$-$1.93          \\
HE~0002$-$1037{$^{\textit{4}}$} &  5010      &  2.00   & 1.80          &$-$2.40          \\
HE~0059$-$6540{$^{\textit{4}}$} &  5040      &  2.10   & 1.80          &$-$2.20          \\
HE~0151$-$6007{$^{\textit{4}}$} &  4350      &  1.00   & 2.10          &$-$2.70          \\
HE~1305$+$0007{$^{\textit{5}}$} &  4750      &  2.00   & 2.00          &$-$2.01          \\

\hline
\end{tabular}

{\textit{1.}} Purandardas {\em et al.} (2019), {\textit{2}}. Allen {\em et al.} (2012), {\textit{3}}. Goswami \& Aoki (2010), {\textit{4}}. Hansen {\em et al.} (2019), {\textit{5}}. Goswami {\em et al.} (2006). \\          
\end{table}
}

{\footnotesize
\begin{table*}
\centering
\caption{\bf{Abundance ratios of neutron-capture elements of the sample stars. }}
\label{tab:abundances}	
\begin{tabular}{lccccccc}
\hline                       
Star name       & [Fe/H]     & [C/Fe]     & [Sr/Fe]   & [Y/Fe]     & [Zr/Fe]    & [Ba/Fe]    &  Ref \\
\hline
CD$-$28~1082    &$-$2.45     &  2.19      & 1.44      &  1.61      &   -        &  2.09      &   1  \\
CS~29503$-$010  &$-$1.70     &  1.65      & 1.13      &  1.09      &  1.26      &  1.81      &   2  \\
CS~29528$-$028  &$-$2.15     &  2.76      & 1.72      &  1.99      &  2.17      &  2.49      &   2  \\
HD~209621       &$-$1.93     &  1.25      & 1.02      &  0.36      &  1.80      &  1.70      &   3  \\
HE~0002$-$1037  &$-$2.40     &  1.90      &$<$1.00    &  0.40      &   -        &  2.00      &   4  \\
HE~0059$-$6540  &$-$2.20     &  1.40      & 1.20      &  0.40      &   -        &  1.70      &   4  \\
HE~0151$-$6007  &$-$2.70     &  1.70      & 1.10      &  0.80      &   -        &  2.30      &   4  \\
HE~1305$+$0007  &$-$2.01     &  1.84      & 0.86      &  0.73      &  2.09      &  2.32      &   5  \\
\hline                       
Star name       & [La/Fe]    & [Ce/Fe]    & [Pr/Fe]   &[Nd/Fe]     & [Sm/Fe]    & [Eu/Fe]    &  Ref \\
\hline
CD$-$28~1082    &  1.55      &  1.97      &  2.30     &  1.99      &  2.29      &  2.07      &   1  \\
CS~29503$-$010  &  2.16      &  2.05      &   -       &  2.31      &  2.34      &  1.69      &   2  \\
CS~29528$-$028  &  2.21      &  2.47      &   -       &  2.54      &   -        &  2.16      &   2  \\
HD~209621       &  2.41      &  2.04      &  2.16     &  1.87      &  1.46      &  1.35      &   3  \\
HE~0002$-$1037  &  2.00      &  1.70      &  2.10     &  2.10      &   -        &  1.70      &   4  \\
HE~0059$-$6540  &  1.60      &  1.40      &  1.40     &  1.70      &   -        &  1.50      &   4  \\
HE~0151$-$6007  &  2.50      &  2.40      &  2.60     &  2.60      &   -        &  2.30      &   4  \\
HE~1305$+$0007  &  2.56      &  2.53      &  2.38     &  2.59      &  2.60      &  1.97      &   5  \\
\hline   
\end{tabular}

References: 1. Purandardas {\em et al.} (2019), 2. Allen {\em et al.} (2012), 3. Goswami \& Aoki (2010), 4. Hansen {\em et al.} (2019), 5. Goswami {\em et al.} (2006).\\          
\end{table*}
}

\section{Parametric-model based study}
\label{sec:parametric_model}
In order to understand the origin of the observed abundances in the sample 
stars, it is important to identify the contribution of the dominant 
neutron-capture process. Following the procedure described  
in Goswami {\em et al.} (2010) and references therein, we have performed 
a parametric-model based study to trace the contributions of s- and 
r-process nucleosynthesis to the observed abundances of the heavy elements.
 We have normalized the Solar s- and r-process isotopic abundances of 
the stellar models of Arlandini {\em et al.} (1999) to the barium 
abundances of the corresponding CEMP-r/s stars. The observed elemental 
abundances of the sample stars are then fitted  with the parametric
model function log $\epsilon_{i}$ = C$_{s}$N$_{is}$ + C$_{r}$N$_{ir}$, 
where N$_{is}$ indicates the normalized abundance  
from s-process, N$_{ir}$ indicates the normalized abundance from 
r-process. C$_{s}$ indicates the component coefficient that corresponds 
to contributions from the s-process and C$_{r}$ indicates the component 
coefficient that corresponds to contributions from the r-process.
    
\begin{figure*}
     \begin{center}
\centering
        {%
                 \label{fig:1}
            \includegraphics[height=8.0cm,width=8.5cm]{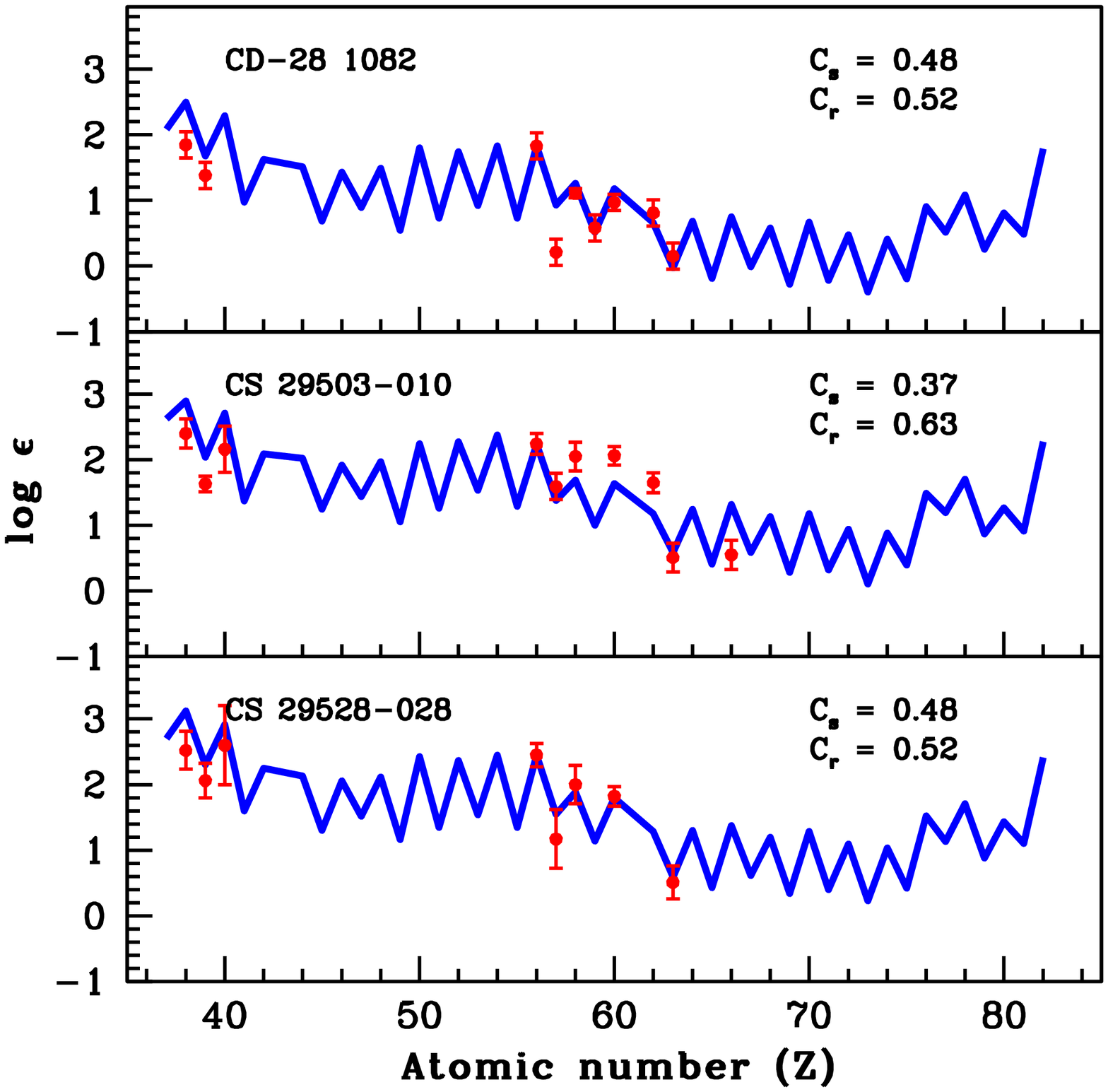}
        }%
        {%
      \label{fig:2}
            \includegraphics[height=8.0cm,width=8.5cm]{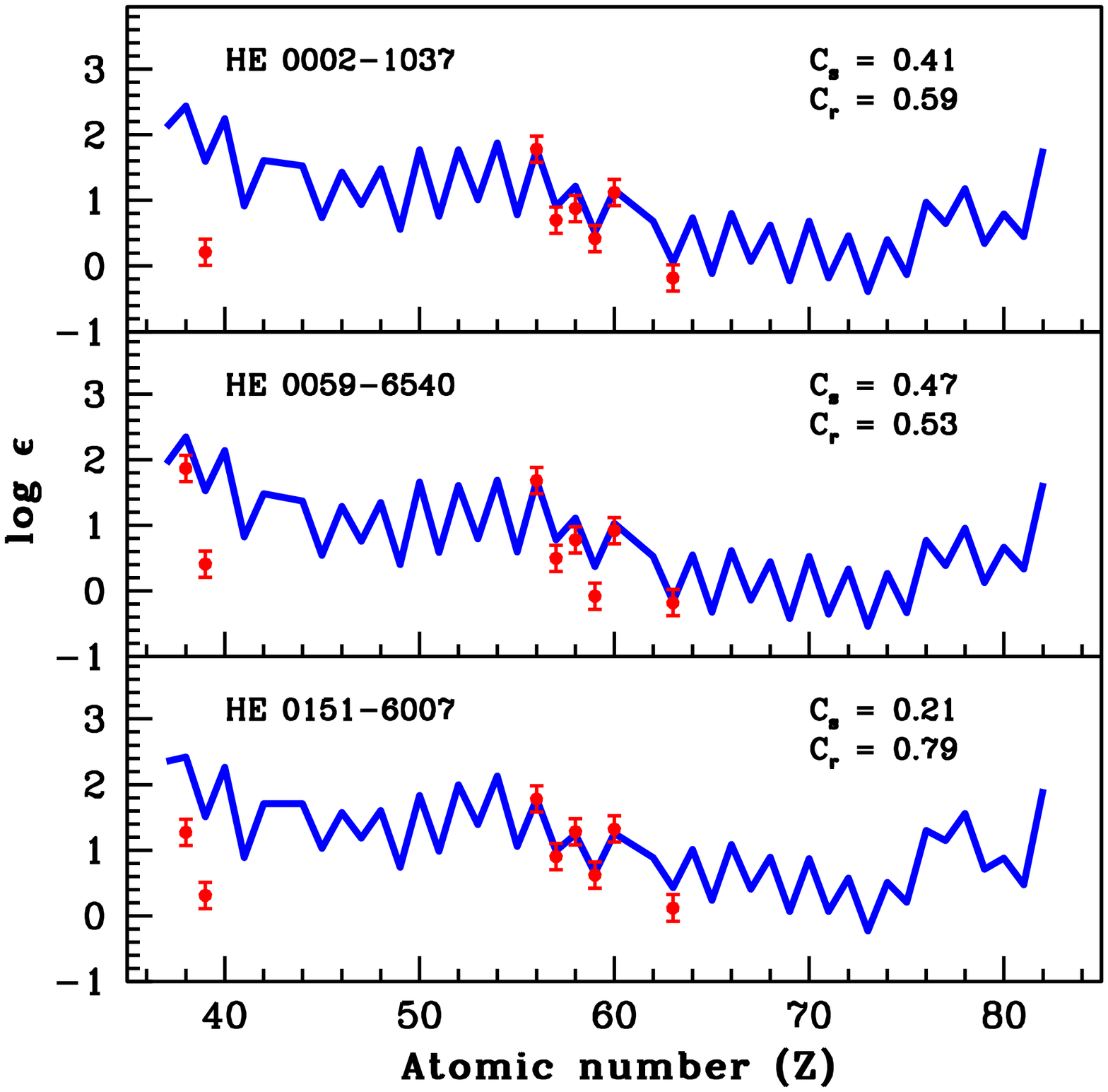}

        }
    \caption{Solid curve represents the best fit for the parametric-model function $log{\epsilon_i}$ = $C_sN_{is}$ + $ C_r N_{ir}$, where $N_{is}$ and $N_{ir}$ represent the abundances due to s- and r-process respectively (isotopic abundances from the Stellar model of Arlandini {\em et al.} (1999) are normalized to the Ba abundances of the corresponding stars). The points with error bars indicate the observed abundances.}
		\label{fig:paramodel_s_r}
       \end{center}
\end{figure*}

The best fitting coefficients and $\chi^{2}$ values are presented 
in Table~\ref{tab:cs&cr}. Figure~\ref{fig:paramodel_s_r} shows the best 
model fits with the observed abundances of the sample stars. 
Goswami \& Aoki (2010) performed a parametric-model based analysis
of the heavy element abundances observed in HD~209621 and HE~1305$+$0007 and 
concluded that similar contributions of both s- and r-process are required 
to explain the abundance pattern of the stars 
(see Figure 5 of Goswami \& Aoki, 2010). WE have  confirmed these
results based on our analysis.

{\footnotesize
\begin{table}
\caption{\bf{Best-fit coefficients and $\chi^{2}$ for the parametric model function $log{\epsilon_i}$ = $C_sN_{is}$ + $ C_r N_{ir}$ }}
\centering
\label{tab:cs&cr}
\begin{tabular}{lccc}
\hline
Star name                        & C$_{s}$          & C$_{r}$          &  $\chi^{2}$   \\
                                 &                  &                  &               \\
\hline
CD$-$28~1082                     &  0.48            &  0.52            &  4.97         \\
CS~29503$-$010                   &  0.37            &  0.63            &  6.76         \\
CS~29528$-$028                   &  0.48            &  0.52            &  1.06         \\
HD~209621{$^{\textit{a}}$}       &  0.57            &  0.52            &  1.80         \\
HE~0002$-$1037                   &  0.41            &  0.59            & 10.73         \\
HE~0059$-$6540                   &  0.47            &  0.53            &  7.83         \\
HE~0151$-$6007                   &  0.21            &  0.79            & 11.91         \\
HE~1305$+$0007{$^{\textit{a}}$}  &  0.47            &  0.53            &  1.07         \\

\hline
\end{tabular}

{\textit{a}}- Goswami \& Aoki (2010).
\end{table}
}

\section{Discussions}
\label{sec:discussion}
\subsection{Different formation scenarios of CEMP-r/s stars}
\label{sec:formation_scenarios}
In order to explain the peculiar abundance pattern of CEMP-r/s stars, 
different authors have proposed several formation scenarios 
(Jonsell {\em et al.} 2006; Lugaro {\em et al.} 2009; Abate {\em et al.} 2016).
 Here, we have discussed briefly  some of the scenarios that are considered
relevant for the sample of CEMP-r/s stars under this study.

(i) Radiative levitation, where the partially ionized heavy elements having 
large photon absorption cross-sections are pushed outwards by radiative 
pressure, could be a possible scenario for the observed over-abundance 
of heavy elements in the CEMP-r/s stars. However,  the simulations of 
Richard {\em et al.} (2002) and Matrozis \& Stancliffe (2016) have shown
 that the process of radiative levitation can occur in the hot stars in 
their main-sequence and main-sequence turn-off phases of evolution due 
to their thin convective envelopes. The sample stars in this study are 
low-temperature objects with temperature in the  range  4350 K $-$ 7100 K,
and the  radiative 
levitation scenario may not be applicable to explain  the observed 
abundance pattern of these stars. This scenario has also been discussed 
at length  by Cohen {\em et al.} (2003), Jonsell {\em et al.} (2006) 
and Abate {\em et al.} (2016) and rejected  as a possible formation mechanism 
of CEMP-r/s stars.

(ii) There are two scenarios where the enhancement of r-process elements 
in the CEMP-r/s stars is attributed to the r-process material enriched 
ISM from which the star is formed. In these scenarios, it is proposed 
that the enrichment of s-process elements is  either due to 
self-contamination in its AGB phase (Hill {\em et al.} 2000; 
Cohen {\em et al.} 2003; Jonsell {\em et al.} 2006) or AGB pollution in 
a binary system (Hill {\em et al.} 2000; Cohen {\em et al.} 2003; 
Jonsell {\em et al.} 2006; Ivans {\em et al.} 2005; Bisterzo {\em et al.} 2011).

In the self-contamination scenario, the star needs to pass through the 
AGB phase of evolution in order to undergo s-process nucleosynthesis. 
However, the evolutionary stage of CEMP-r/s stars  studied so far  
reveals that they have not yet passed through the Red 
Giant Branch phase to produce the s-process elements. For this reason, this 
scenario had been rejected by many authors (Jonsell {\em et al.} 2006, 
 Abate {\em et al.} 2016). Two objects CS~29503$-$010 and CS~29528$-$028 
in our sample,  exhibit log g values ($\sim$ 4.0), similar to that of 
dwarfs or sub-giants. The rest of the stars show low values 
($\sim$ 1 $-$ 24) of $^{12}$C/$^{13}$C implying the extrinsic nature of 
heavy-elements and carbon in these stars. Thus, this scenario is not 
applicable to our sample  stars.

In another scenario, the primary, being more massive than the other in the 
binary system formed from the r-process material enriched ISM, evolves 
faster and proceeds through the AGB phase producing s-process elements 
along with carbon. The mass transfer episodes in the AGB phase then make 
the secondary star enriched in s-process elements. Considering this 
scenario of pre-enrichment of the binary system with r-process elements 
from the r-rich molecular cloud, Bisterzo {\em et al.} (2011, (2012) tried 
to reproduce the observed [hs/ls] in the CEMP-r/s stars. Although they 
claimed to be successful in doing so (compatible within error bars), there 
are still some arguments against this scenario. It is found that the 
abundances of Ba and Eu correlate in CEMP-r/s stars and the AGB models 
cannot explain this correlation in case of independent enrichment of 
s- and r-process elements (Abate {\em et al.} 2016). Also, the large 
fraction of CEMP-r/s stars among the CEMP-s stars cannot be explained 
by this scenario (Jonsell {\em et al.} 2006; Lugaro {\em et al.} 2009).

(iii) This scenario explains how a star can acquire s- and r-process 
elements in a triple star system. The most massive one, among the three 
stars, evolves the fastest and supernova explosion of this star makes 
the other two stars r-rich. Then, the more massive one among the other 
two stars evolves through the AGB phase and produces s-process elements 
along with carbon. AGB Mass-transfer from this star makes the tertiary 
a CEMP-r/s star (Cohen {\em et al.} 2003;
Jonsell {\em et al.} 2006). However,  it seems very 
unlikely that after the SN explosion the triple system survives for 
further mass transfer. Abate {\em et al.} (2016)  couldn't reproduce the 
observed frequency of CEMP-r/s stars among CEMP-s stars and hence 
dismissed this scenario. 

(iv) There are two proposed scenarios that considered binary systems where 
the s-process elements in the CEMP-r/s star are assumed to come from 
the primary through its AGB phase and r-process elements are attributed 
to either Type 1.5 supernova (Jonsell {\em et al.} 2006; Zijlstra 2004) 
or an accretion-induced collapse (AIC) (Qian \& Wasserburg 2003; 
Cohen {\em et al.} 2003).

In the first scenario, the primary star, being more massive than the other 
in the binary system, evolves through the AGB phase. During the AGB phase, 
the star produces and transfers s-process rich material to the companion. 
Then, the AGB star may explode as a Type 1.5 supernova and pollute the 
companion with r-process elements making the secondary a CEMP-r/s star. 
Iben \& Renzini (1983) gave the name `Type 1.5 supernovae' to the process 
when the degenerate cores of high mass AGB stars, due to low mass-loss 
efficiency at low-metallicity, remain as massive as to reach Chandrasekhar 
mass limit and explode (Zijlstra 2004). However, Nomoto {\em et al.} (1976), 
Iben \& Renzini (1983) and Lau {\em et al.} (2008) stated that Type 1.5 
supernova can destroy the primary star and hence disrupt the binary system. 
As most of the CEMP-r/s stars are reported to be found in binary systems 
(Lucatello {\em et al.} 2005) Abate {\em et al.} (2016) rejected this 
scenario.

In the other scenario, after transferring the AGB processed material 
(s-process elements) to the companion star, the primary star becomes a 
white dwarf. Then, as time progresses the secondary star evolves to giant 
branch and transfers material back to the white-dwarf. This mass-transfer 
may trigger an accretion-induced collapse (AIC) and hence pollute the 
secondary star with r-process material (Qian \& Wasserburg 2003; 
Cohen {\em et al.} 2003). This scenario demands the secondary star to be 
in the giant branch to fill the Roche-lobe for the second phase of 
mass-transfer. This scenario may be rejected as the 
observed CEMP-r/s stars are not always giants (Lugaro {\em et al.} (2009)). 
In many cases, these stars 
are seen in the main-sequence turn-off making the accretion process 
difficult. Abate {\em et al.} (2016) stated that the three phases of 
mass-transfer work properly only if the orbital separation of the binary 
system is narrow, but, the observed frequency of CEMP-r/s stars could not 
 be reproduced even by  considering  a narrow separation. Also, there are 
uncertainties regarding the efficiency of AIC to produce enough r-process 
elements to match the observed abundances of heavy-elements in CEMP-r/s 
stars (Qian \& Woosley 1996; Qian \& Wasserburg 2003).

(v) A  formation scenario named ``intermediate neutron-capture process 
or i-process", similar to that of the formation scenarios of CH, Ba and 
CEMP-s stars has been proposed recently to explain the abundances of 
CEMP-r/s stars. This scenario considers a binary system, where one star 
is slightly massive than the other. The more massive star evolves faster and passes through the AGB phase polluting the companion with AGB processed material. The difference of this scenario with that of the s-process enrichment scenarios is that in this scenario a neutron density (n $\sim$ 10$^{15}$ cm$^{-3}$), which is intermediate to the neutron-densities of both s- and r-process, can produce the double enhancement seen in CEMP-r/s stars (Cowan \& Rose 1977; Dardelet {\em et al.} 2014; Hampel {\em et al.} 2016). Cowan \& Rose (1977), for the first-time, suggested the possibility of occurrence of i-process in AGB stars. They found that significantly high neutron-density can be achieved by mixing hydrogen-rich material into the intershell region of AGB stars. Using i-process models Dardelet {\em et al.} (2014) and Hampel {\em et al.} (2016) have successfully reproduced the abundance distribution of a number of well-known CEMP-r/s stars.

\subsection{Comparison of the abundance pattern of the sample stars with i-process model}
\label{sec:i_process}
Dardelet {\em et al.} (2014) and Hampel {\em et al.} (2016) calculated i-process model yields with slightly different approaches, but, both the groups could successfully reproduce the observed abundance pattern of CEMP-r/s stars. Single-zone nuclear network calculations were used in both the studies. Assuming proton-ingestion from the H-rich envelope to the He pulse-driven convective zone (PDCZ) to be responsible for the generation of higher neutron-densities (n $\sim$ 10$^{15}$ cm$^{-3}$) Dardelet {\em et al.} (2014) used a constant combined mass-fraction of C+H (= 0.7) in their simulations. They considered the termination time of the i-process as a free parameter for their calculations. On the other hand, Hampel {\em et al.} (2016)  calculated the yields of the neutron-capture nucleosynthesis, assuming the nucleosynthesis to operate in the intershell region of an AGB star, at different constant neutron-densities starting from n $\sim$ $10^{7}$ cm$^{-3}$ to $10^{15}$ cm$^{-3}$. Dardelet {\em et al.} (2014) assumed the temperature and density for the He PDCZ to be $2.0 \times 10^{8}$ K and 10$^{4}$ gcm$^{-3}$ respectively. These physical input parameters are chosen so as to prevent the proton capture by $^{13}$N and allow the $^{13}$C($\alpha$, n)$^{16}$O reaction for neutron-release. As a test, Hampel {\em et al.} (2016) tried to calculate the yields with a range of temperatures ($1 \times 10^{8}$ K to $2.2 \times 10^{8}$ K) and densities (800 gcm$^{-3}$ to 3200 gcm$^{-3}$), but, didn't see significant changes in the results. However, for the final simulations they adapted the parameters (T = $1.5 \times 10^{8}$ K and $\rho$ = 1600 gcm$^{-3}$) of the intershell region of a low-metallicity (z = 10$^{-4}$), low-mass (M = 1 M$_{\odot}$) AGB star (Stancliffe {\em et al.} 2011). As initial abundances of the He PDCZ, Dardelet {\em et al.} (2014) considered solar-abundances (except C and O), scaled down to z = 10$^{-3}$. The abundances of C and O are taken to be X($^{12}$C)=0.5 and X($^{16}$O)=0.05, which are typical abundances for the He PDCZ. Whereas Hampel {\em et al.} (2016) adapted the constituents of the intershell region from that of Abate {\em et al.} (2015). A high neutron exposure of $\tau$ $\sim$ 495 mb$^{-1}$ is ensured by adjusting the run times of the models. Due to such high neutron exposures, the abundance pattern of heavy elements and the seed nuclei comes to an equilibrium, which makes the element-to-element ratio a function of constant neutron-density.

In the model of Dardelet {\em et al.} (2014), almost all the isotopes of $^{12}$C get transformed into $^{13}$N during the first second of run-time. Then in 9.97 minutes, $^{13}$N decays to form $^{13}$C, which captures $\alpha$ to release neutrons with high neutron-densities through the reaction $^{13}$C($\alpha$, n)$^{16}$O. The neutron-exposure ($\tau$) increases with time reaching up to 10 $-$ 50 mb$^{-1}$ and subsequently the heavier elements are produced. This model could successfully reproduce the observed abundance pattern of three CEMP-r/s 
stars. On the other hand, Hampel {\em et al.} (2016) noticed that when the neutron 
exposure was kept switched on, for lower neutron densities (n $\sim$ $10^{7}$ cm$^{-3}$), typical s-process abundance pattern is produced with stable peaks of ls (Sr, Y, Zr) and hs (Ba, La, Ce) elements. But with higher neutron-densities (n = $10^{12}-10^{15}$ cm$^{-3}$), both the peaks of ls and hs elements shift to lighter elements. In particular, a peak at $^{135}$I is formed due to the i-process neutron-densities. Then, the neutron exposure is turned off for t = 10 Myr. During this time, it is noticed that unstable isotopes decay to produce stable isotopes at ls and hs peaks. The decay of $^{135}$I produces $^{135}$Ba. With increasing neutron-densities abundances of Ba and Eu are found to increase. This is how the abundance pattern gets modified due to i-process. Using this i-process model, Hampel {\em et al.} (2016) could successfully reproduce the observed abundance pattern of twenty CEMP-r/s stars, including the three previously reproduced by the i-process model of Dardelet {\em et al.} (2014).
 
We have used the model predictions ([X/Fe]) for neutron densities ranging 
from n $\sim$ $10^{9}-10^{15}$ cm$^{-3}$ ( Hampel {\em et al.} (2016)),  
and compared with the elemental abundance pattern of our sample of 
CEMP-r/s stars. In order to examine whether the i-process models could 
reproduce the observed abundances of the sample stars we have followed 
the procedure discussed in Hampel {\em et al.} (2016) and used the equation-
\begin{equation}
X = X_{i} \times (1-d) + X_{\odot}\times d
\end{equation}
where $X_{i}$ is the model yield, $X_{\odot}$ is solar-scaled abundance 
and $d$ is a dilution factor. 

{\footnotesize
\begin{table}
\caption{\bf{Fit parameters of i-process model for the sample stars}}
\centering
\label{tab:i-process_fit_parameters}
\begin{tabular}{lccc}
\hline
Star name                        &  Neutron-density  &   d              &  $\chi^{2}$   \\
                                 &   n (cm$^{-3}$)   &                  &               \\
\hline
CD$-$28~1082                     &  10$^{13}$        &  0.9704          &  3.26         \\
CS~29503$-$010                   &  10$^{13}$        &  0.9745          &  2.74         \\
CS~29528$-$028                   &  10$^{12}$        &  0.9331          &  0.51         \\
HD~209621                        &  10$^{13}$        &  0.9798          &  3.52         \\
HE~0002$-$1037                   &  10$^{14}$        &  0.9766          &  1.71         \\
HE~0059$-$6540                   &  10$^{13}$        &  0.9908          &  2.34         \\
HE~0151$-$6007                   &  10$^{14}$        &  0.9274          &  2.06         \\
HE~1305$+$0007                   &  10$^{14}$        &  0.9262          &  2.62         \\

\hline
\end{tabular}

\end{table}
}

\begin{figure*}
     \begin{center}
\centering
        {%
                 \label{fig:1}
            \includegraphics[height=8.0cm,width=8.5cm]{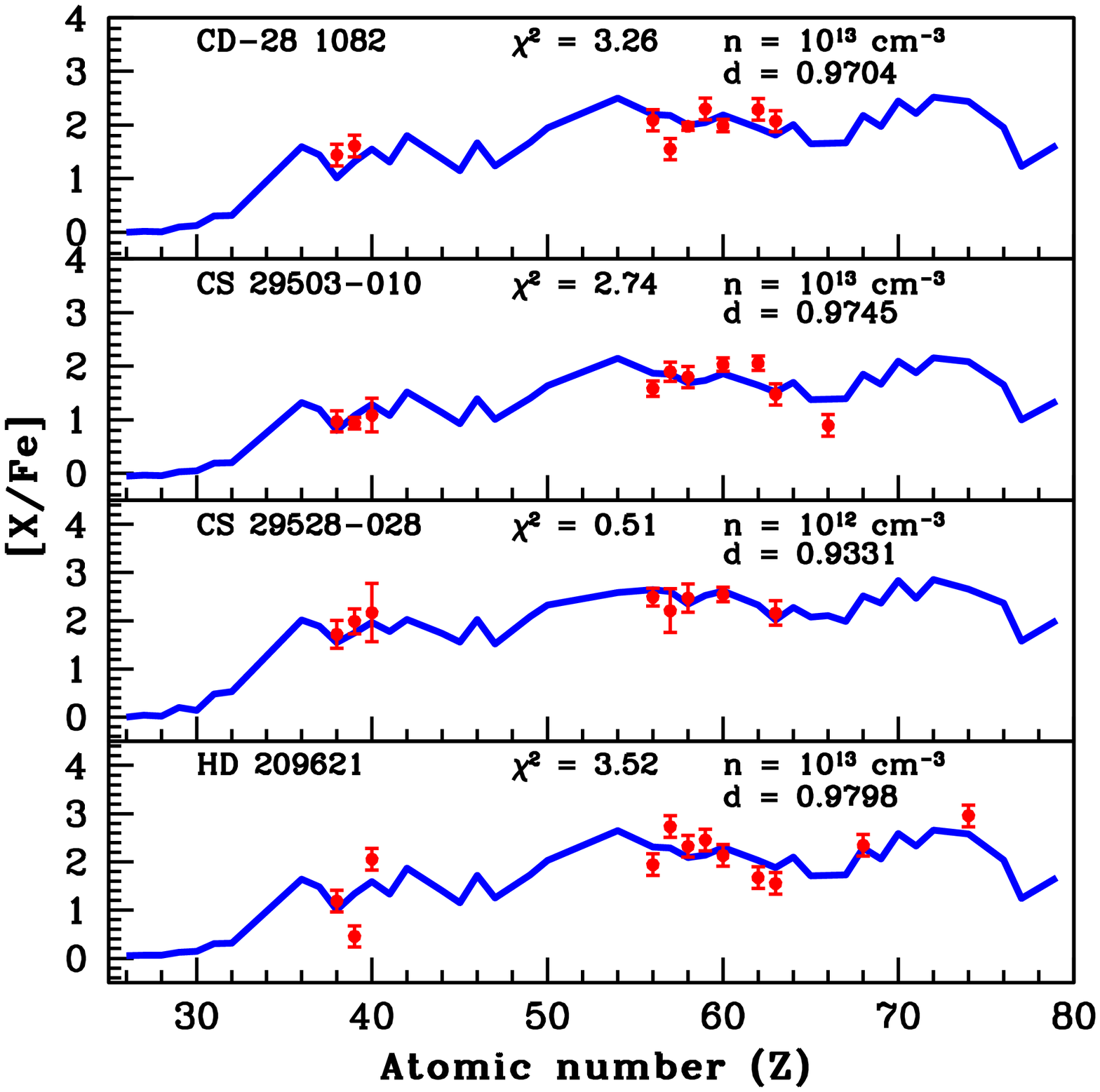}
        }%
        {%
      \label{fig:2}
            \includegraphics[height=8.0cm,width=8.5cm]{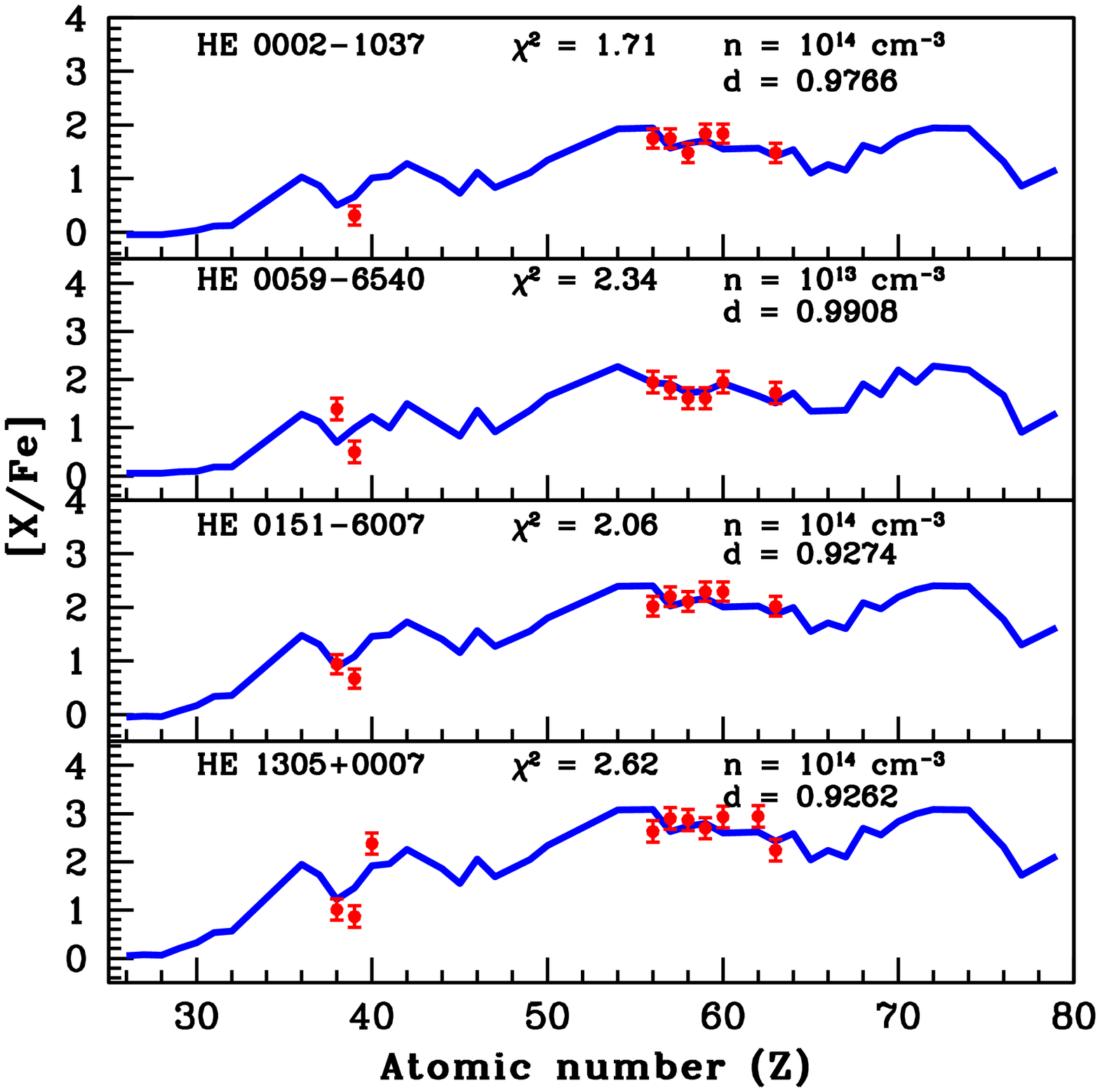}

        }
    \caption{Best-fitting i-process model (solid blue curve) for the sample stars. The points with error bars indicate the observed abundances.}
   \label{fig:iprocess}
       \end{center}
\end{figure*}

Table~\ref{tab:i-process_fit_parameters} presents the values of the fit 
parameters neutron-density, `d' and `$\chi^{2}$' for each star. The best 
fit neutron-density for each of the sample stars is chosen for which we 
got the minimum value of `$\chi^{2}$'. The best fit models with 
appropriate neutron-densities and corresponding dilution factors are shown 
in Figure~\ref{fig:iprocess}. We have found that i-process models with 
neutron-densities of n $\sim$ $10^{12}$ cm$^{-3}$ to $10^{14}$ cm$^{-3}$ 
closely fit the observed abundances of the sample stars. We have obtained
 a neutron-density of n $\sim$ 10$^{14}$ cm$^{-3}$ for the star HE~1305$+$0007,
 which is same as reported by Hampel {\em et al.} (2016) for the object. 
Hampel {\em et al.} (2019) tried to understand the i-process nucleosynthesis 
through the abundances of Pb in the CEMP-r/s stars. 
In their i-process models Hampel {\em et al.} (2019) considered 
neutron-exposure ($\tau$) as a free parameter along with dilution 
factor (d). Varying `d' and `$\tau$' at different constant neutron-densities, 
they could fit the observed abundances of HE~1305+0007 and HD~209621 with 
models of neutron-densities, n $\sim$ 10$^{14}$ cm$^{-3}$ 
and 10$^{13}$ cm$^{-3}$ respectively. We, too, got the best fit 
at n $\sim$ 10$^{13}$ cm$^{-3}$ for the object HD~209621 taking only `d' as 
a free parameter with constant `$\tau$'.

\section{Conclusions}
\label{sec:conclusion}
In this study, we have examined if the currently available theoretical
i-process stellar yields could adequately explain the enhancement
in both s- and r-process elements observed in a selected sample of  
CEMP-r/s stars. We have considered eight stars, out of which, the 
abundances of the  heavy elements for three objects
are taken from our previous studies (Goswami et al. 2006, Goswami \&
Aoki 2010, Purandardas et al. 2019) and  the rest from other sources
(Allen et al. 2012,  Hansen et al. 2019).

With the help of a parametric-model based study, we have estimated the 
contributions from s- and r-process nucleosynthesis to the observed 
elemental abundances. It is  found that  similar 
contributions  from both the processes are required to explain the 
observed abundance pattern of heavy-elements in these stars. 

With reference to the double enhancement seen in the sample stars, we 
have discussed different formation scenarios of CEMP-r/s stars. The 
scenarios involving two stellar sites for the production of s- and r-process 
elements are found to be not applicable to the sample of CEMP-r/s stars 
under this study. However, we have seen that i-process models of 
Hampel {\em et al.} (2016) can satisfactorily reproduce the observed 
overabundance of heavy elements in these stars. The i-process stellar yields
required to fit the observed abundance patterns are found
to correspond to neutron-densities as high
as $10^{12}$ cm$^{-3}$ to $10^{14}$ cm$^{-3}$.
The estimated   low values of $^{12}$C/$^{13}$C ratio observed in the stars 
agree well with the i-process predictions indicating the extrinsic nature 
of carbon and  the heavy elements.


\section*{Acknowledgements}

The funding from the  DST SERB project  EMR/2016/005283  is gratefully 
acknowledged.  We are thankful to Melanie Hampel for providing us with the  
i-process yields in the form of number fractions. This work made use of 
the SIMBAD astronomical database, operated at CDS, Strasbourg, France 
and the NASA ADS, USA.

\end{document}